\documentclass[twocolumn, twocolappendix]{aastex631}
\usepackage{float}
\usepackage{wasysym}
\usepackage{booktabs}
\usepackage[encapsulated]{CJK}

\begin{document}
\begin{abstract}
The first JWST data revealed an unexpected population of red galaxies
that appear to have redshifts of $z\sim 7-9$ and
high masses of $M_*\gtrsim 10^{10}\mathrm{M}_{\astrosun}$.
Here we fit  S\'{e}rsic profiles to the F200W NIRCam images of the
13 massive galaxy candidates of Labb\'e et al.,
to determine their structural parameters.
Satisfactory fits were obtained for nine galaxies.
We find that their effective radii are extremely small, ranging from $r_{\rm e}\sim 80$\,pc to $r_{\rm e} \sim 300$\,pc, with a mean of
$\langle r_{\rm e}\rangle \approx 150$\,pc.
For their apparent stellar masses, the galaxies are smaller
than any other galaxy population that has been observed at any other redshift.
We use the fits to derive circularized three-dimensional stellar mass profiles of the galaxies, and compare these to the mass profiles of massive
quiescent galaxies at $z\sim 2.3$ and nearby elliptical galaxies. 
Despite the fact that the high redshift galaxies have $10-20$ times smaller half-light radii than their putative descendants, 
the central stellar densities are very similar.
The most straightforward
interpretation is that the dense compact inner regions of the most massive ellipticals today were already in place $\sim 600$\,Myr after the Big Bang. We caution that the
redshifts and masses of the galaxies
remain to be confirmed, and that the
complex NIRCam point spread function is
not yet fully characterized.
\end{abstract}

\title{Sizes and mass profiles of candidate massive galaxies discovered by JWST at $7<z<9$:\\ evidence for very early formation of the central $\sim 100$\,pc of present-day ellipticals}
\author[0009-0005-2295-7246]{Josephine F.W. Baggen}
\affiliation{Department of Astronomy, Yale University, New Haven, CT 06511, USA}
\author[0000-0002-8282-9888]{Pieter van Dokkum}
\affiliation{Department of Astronomy, Yale University, New Haven, CT 06511, USA}
\author[0000-0002-2057-5376]{Ivo Labb\'e}
\affiliation{Centre for Astrophysics and Supercomputing, Swinburne University of Technology,
Melbourne, VIC 3122, Australia}
\author[0000-0003-2680-005X]{Gabriel Brammer}
\affiliation{Cosmic Dawn Center (DAWN), Niels Bohr Institute, University of Copenhagen, Jagtvej 128,
K{\o}benhavn N, DK-2200, Denmark}

\author[0000-0001-8367-6265]{Tim B. Miller}
\affiliation{Department of Astronomy, Yale University, New Haven, CT 06511, USA}
\author[0000-0001-5063-8254]{Rachel Bezanson}
\affiliation{Department of Physics and Astronomy and PITT PACC, University of Pittsburgh, Pittsburgh,
PA 15260, USA}

\author[0000-0001-6755-1315]{Joel Leja}
\affiliation{Department of Astronomy \& Astrophysics, The Pennsylvania State University, University Park, PA 16802, USA}
\affiliation{Institute for Computational \& Data Sciences, The Pennsylvania State University, University Park, PA 16802, USA}
\affiliation{Institute for Gravitation and the Cosmos, The Pennsylvania State University, University Park, PA 16802, USA}
\author[0000-0001-9269-5046]{Bingjie Wang (\begin{CJK*}{UTF8}{gbsn}王冰洁\ignorespacesafterend\end{CJK*})}
\affiliation{Department of Astronomy \& Astrophysics, The Pennsylvania State University, University Park, PA 16802, USA}
\affiliation{Institute for Computational \& Data Sciences, The Pennsylvania State University, University Park, PA 16802, USA}
\affiliation{Institute for Gravitation and the Cosmos, The Pennsylvania State University, University Park, PA 16802, USA}
\author[0000-0001-7160-3632]{Katherine E. Whitaker}
\affiliation{Department of Astronomy, University of Massachusetts, Amherst, MA 01003, USA}
\affiliation{Cosmic Dawn Center (DAWN), Niels Bohr Institute, University of Copenhagen, Jagtvej 128,
K{\o}benhavn N, DK-2200, Denmark}
\author[0000-0002-1714-1905]{Katherine A. Suess}
\affiliation{Department of Astronomy and Astrophysics, University of California, Santa Cruz, 1156 High Street, Santa Cruz, CA 95064 USA}
\affiliation{Kavli Institute for Particle Astrophysics and Cosmology and Department of Physics, Stanford University, Stanford, CA 94305, USA}
\author[0000-0002-7524-374X]{Erica J. Nelson}
\affiliation{Department for Astrophysical and Planetary Science, University of Colorado, Boulder, CO, USA}

\section{Introduction}
Our first glimpse into the $z\sim$11 universe came from detections of young, star-forming galaxies with the combined power of the Hubble and Spitzer Space Telescopes \citep[e.g.][]{Oesch2016}.
The physical properties of galaxies at the redshift frontier were well characterized by relatively low stellar masses,
with virtually no galaxies with $M_* \gtrsim 10^{10}$ detected at $z\gtrsim 6$ \citep[see, e.g.,][]{Stefanon2021}.  However, 
the landscape is rapidly changing, as the James Webb Space Telescope (JWST) enables us to detect galaxies that are fainter, redder, and at higher redshifts \citep[e.g.,][]{Castellano2022,Labbe2023,Finkelstein2022,Atek2023,Austin2023, Boyett2023, Donnan2023, Looser2023, Mason2023, Naidu2022a, Naidu2022b}. 

In particular, \citet{Labbe2023} (L23 hereafter) discovered 13 massive galaxy candidates at $z = 6.5-9.1$ with two pronounced breaks in their spectral energy distributions (SEDs) in the Cosmic Evolution Early
Release Science (CEERS) survey \citep{Finkelstein2022, Finkelstein2023}. 
As discussed in \citet{BoylanKolchin2023} these objects may
pose a challenge for galaxy formation theory, and perhaps even for the $\Lambda$CDM
model, if the masses and redshifts are correct. The halo mass function provides an upper limit on the stellar mass that can form at a given redshift,
as the stellar mass cannot exceed the amount of baryons within the halo at a given time: $M_{*} \leq \epsilon f_b M_{\mathrm{halo}}$, with $f_b$ the cosmic baryon fraction $f_b = \Omega_b/\Omega_m$ and  $\epsilon\leq1$ being the efficiency of converting baryons into stars. Typical values for the efficiency of baryon conversion ($\epsilon$=0.1 and $\epsilon$=0.32) imply a
too high stellar mass density than allowed by $\Lambda$CDM for these redshifts. The only way to be consistent with the data 
is to require extreme efficiencies  converting baryons into stars ($\epsilon\geq$0.57-0.99) in the early Universe.  Such high star formation efficiencies have never been observed directly, though there is indirect evidence through the modeled star formation histories of the first quiescent galaxies (Antwi-Donso et al. in prep).

Given the implications of a population of early massive galaxies, there have been several studies proposing alternate solutions to solve this tension. 
The stellar masses may be overestimated, with less extreme estimates when adopting different star formation histories, different initial mass functions \citep{Haslbauer2022}, different synthetic templates \citep{Steinhardt2022}, or taking into account extreme emission lines \citep{Endsley2023}.
Other more exotic solutions include different dark energy models \citep{Menci2022}, Early Dark Energy \citep{BoylanKolchin2023}, 
fuzzy dark matter \citep{Gong2023}, primordial black holes or axion mini clusters \citep{Hutsi2023}, or cosmic strings \citep{Jiao2023}.  true nature remains a puzzle that needs to be solved. 
Finally, contamination by red AGN is also possible and perhaps likely, supported by the detection of a population of compact red sources in JWST imaging data up to $z\sim8$ \citep[e.g.][]{Furtak2022a, Fujimoto2023, Ono2022,Barro2023, Onoue2023, Labbe2023b, Oesch2023} and through broad line identification with JWST spectroscopy up to $z\sim7$ \citep[e.g.][]{Kocevski2023, Harikane2023, Matthee2023}.

Here we take the stellar masses and redshifts of the L23 sample at face value and ask what the structure is of the sources.
Their sizes and morphologies provide clues to their formation and, by extension, to the evolution of the most massive
galaxies in the universe. Furthermore, we can test whether most or all of the candidates are dominated by the light of active nuclei, 
as the objects should then be point sources in the JWST imaging. 
Throughout this work we assume $\Lambda$CDM cosmology with $H_0$=70 km/s/Mpc, $\Omega_{m,0}$=0.3 and $\Omega_{\Lambda,0}$=0.7.

 \begin{figure*}
\begin{center}
\hspace{-1.2cm}
\includegraphics[width=\textwidth, trim={0.2cm, 2cm, 0cm, 1cm},clip]{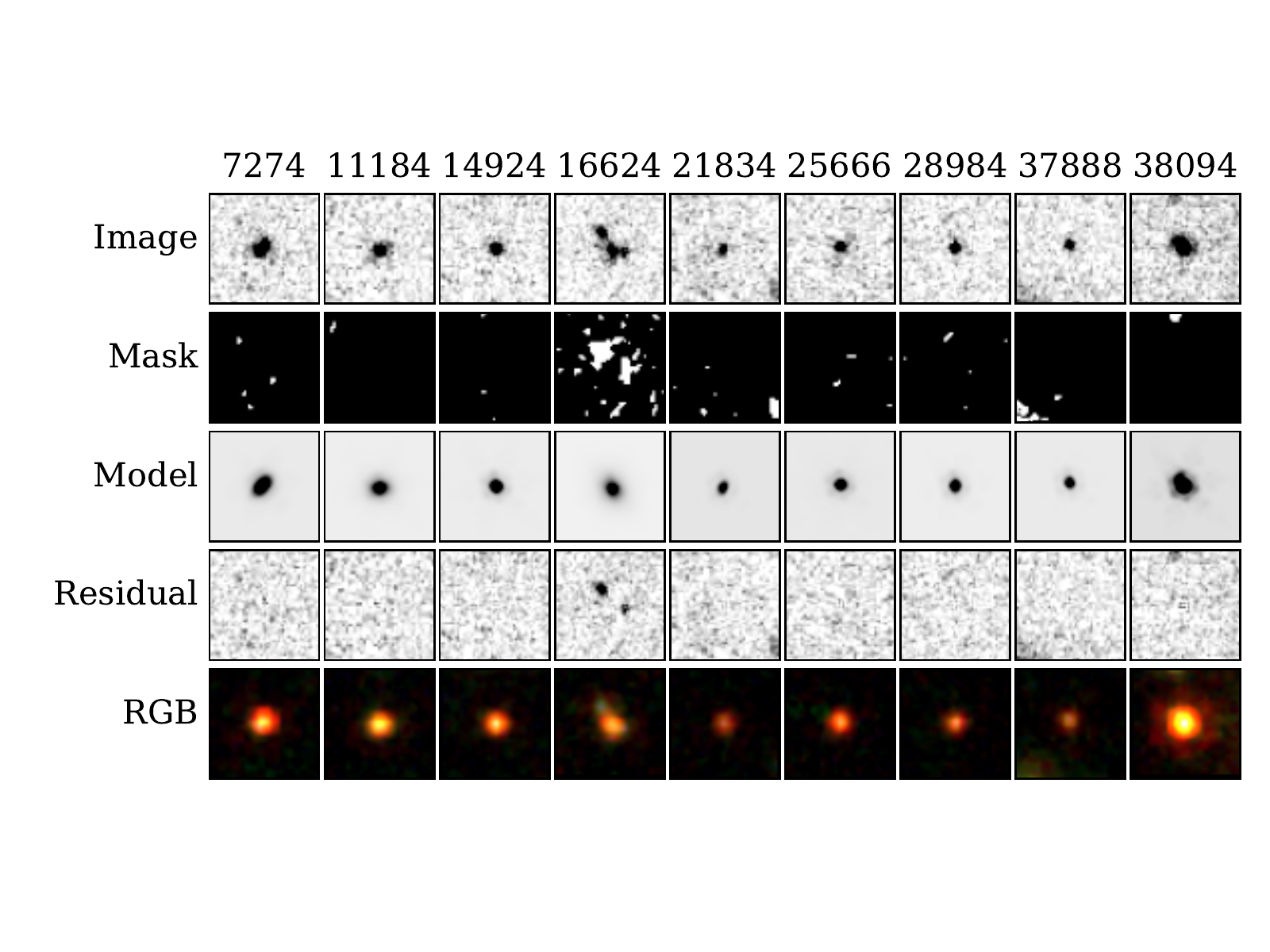}
\end{center}
  \caption{ 
  The top row shows $1\arcsec \times 1\arcsec$ images in the NIRCam F200W band of the nine galaxies for which a good S\'ersic fit was found. The second row shows the masks that were used, the third the best-fitting \textsc{galfit} models. The residuals, obtained by subtracting the best-fitting models from the images, are shown in the fourth row.
   The bottom row shows RGB images of the galaxies, with F150W as the blue band, F277W as the green band and F444W as the red band. 
   }
     \label{fig:setup_segmentation}
 \end{figure*}

\section{Data}
\label{sec:data}
This paper is based on one of the first data sets that were obtained with the JWST
Near Infrared Camera (NIRCam), the Cosmic Evolution Early Release Science (CEERS) program \citep[PI: Finkelstein; PID: 1345][]{Finkelstein2022, Finkelstein2023} \citep[see data DOI:][]{doiCEERSmosaic}.
The initial data consist of four pointings, covering $\approx 40$\,arcmin$^2$ and overlapping with existing HST fields. The JWST data were taken in six broadband filters, F115W, F200W, F150W, F277W, F356W, and F444W, and a medium bandwidth filter F410M. The images are reduced with the standard JWST calibration pipeline (v1.5.2) followed by the grizli pipeline \citep{gabe_brammer_grizli}.
The final mosaics for each of these bands are available online \citep{mosaicsV4}.
The resolution of these mosaics is $0\farcs 02$\,pix$^{-1}$
for F115W, F150W and F200W and $0\farcs 04$\,pix$^{-1}$
for F277W, F356W and F444W. 

The selection of the 13 candidate massive galaxies is described in detail in L23\footnote{\dataset[https://github.com/ivolabbe/red-massive-candidates]{https://github.com/ivolabbe/red-massive-candidates}}. Briefly,
they all have a ``double-break''  in their spectral energy distributions (SEDs), identified as the Lyman break in the rest-frame UV and the Balmer break in the rest-frame optical. The methodology selects galaxies with high mass-to-light ratios while ensuring that the photometric redshifts are well-constrained. The photometric redshifts and stellar masses were found using three codes, EAZY \citep{Brammer2008}, 
 the Prospector-beta settings \citep{Wang2023} in Prospector \citep{Johnson2021}
 and five configurations of Bagpipes \citep{Carnall2018}.
L23 adopted a Salpeter \citep{Salpeter1955} initial mass function (IMF).
From the 7 different measurements for each galaxy, the median value for the photometric redshift and stellar mass is used in this work, following L23. 
Errors in the stellar masses are from the 16th and 84th median posterior distribution. 
For more details on the sample selection and SED fitting procedure, we refer to the Methods section in L23.

Follow-up spectroscopy is needed to confirm the stellar masses and the high redshift nature of the sources. For 3 of the 13 galaxies in L23 spectra have been obtained with NIRSpec on JWST. One of the sources, L23-13050, is an AGN at $z = 5.624$, as shown in \citet{Kocevski2023} (CEERS 3210 therein).
It was already suspected to be a potential strong-line emitter at $z = 5.72$ by \citet{PerezGonzalez2023}. 
L23-35300 and L23-39575 are confirmed to be at high redshift ($z\sim 8$), as shown in \citet{Fujimoto2023} (CEERS3-1748 and CEERS1-3910). 

We obtain point-spread functions (PSFs) for all the galaxies
using the WebbPSF tool \citep{Perrin2014}. Synthetic PSFs have the advantage of perfect sampling, centroids, and high S/N ratio. However, several studies have found that WebbPSF profiles are too narrow in the core when comparing them to bright point source profiles
\citep{Ding2022, Ono2022, Onoue2023, Weaver2023}.

We therefore also use an empirical PSF, made from stacking isolated, unsaturated stars in the CEERS mosaic. Only stars whose centroid is close to the center of a pixel are considered, and small shifts are applied to center them precisely prior to stacking.
In the analysis we give equal weight to the synthetic and empirical PSF.
We note that a comparison of the radial profiles of stars and the WebbPSF model showed that they are nearly identical in the F200W (which we use for the galaxy fits).

\begin{table*}[t!]
\small{
\begin{center}
\begin{tabular}{lll|lllllllll|ll}

\multicolumn{1}{l}{ID} & \multicolumn{1}{l}{$z_{\mathrm{phot}}$} & $\log$($M_*$/M$_\odot$)   &  $m_{\mathrm{L23}}$ & $m_{\mathrm{webb}}$& $m_{\mathrm{star}}$ & $n_{\mathrm{webb}}$   &  $n_{\mathrm{star}}$ & $r_{\mathrm{e,webb}}$ & $r_{\mathrm{e,star}}$  & $(b/a)_{\mathrm{webb}}$ & $(b/a)_{\mathrm{star}}$ & $n$ & $r_\mathrm{e}$\\
\hline
7274 & 7.77 & 9.87 & 27.1 & 27.2 & 27.1 & 0.5 & 0.5 & 274 & 268 & 0.5 & 0.5 & 0.5$^{+0.4}_{-0.0}$ & 273$^{+18}_{-22}$ \\
11184 & 7.32 & 10.18 & 27.1 & 27.0 & 27.0 & 1.5 & 1.3 & 236 & 207 & 0.7 & 0.8 & 1.4$^{+0.3}_{-0.4}$ & 216$^{+18}_{-32}$ \\
14924 & 8.83 & 10.02 & 27.5 & 27.5 & 27.4 & 1.1 & 0.5 & 96 & 101 & 0.7 & 0.8 & 0.7$^{+0.5}_{-0.2}$ & 108$^{+26}_{-10}$ \\
16624 & 8.52 & 9.30 & 27.0 & 27.1 & 27.1 & 2.2 & 2.0 & 378 & 331 & 0.7 & 0.7 & 1.6$^{+0.6}_{-0.4}$ & 311$^{+85}_{-54}$ \\
21834 & 8.54 & 9.61 & 28.4 & 28.4 & 28.3 & 0.5 & 0.5 & 146 & 138 & 0.1 & 0.1 & 0.5$^{+1.4}_{-0.0}$ & 137$^{+51}_{-60}$ \\
25666 & 7.93 & 9.52 & 27.7 & 27.8 & 27.7 & 4.1 & 1.7 & 80 & 78 & 0.8 & 0.8 & 1.8$^{+4.3}_{-0.7}$ & 86$^{+20}_{-12}$ \\
28984 & 7.54 & 9.57 & 27.9 & 27.9 & 27.8 & 1.5 & 0.5 & 105 & 121 & 0.5 & 0.6 & 0.8$^{+0.7}_{-0.3}$ & 111$^{+12}_{-17}$ \\
37888 & 6.51 & 9.23 & 28.1 & 28.3 & 28.2 & 1.0 & 0.7 & 110 & 108 & 0.3 & 0.4 & 1.3$^{+1.9}_{-0.8}$ & 88$^{+40}_{-27}$ \\
38094 & 7.48 & 10.89 & 26.3 & 26.6 & 26.5 & 1.1 & 0.5 & 79 & 90 & 0.6 & 0.6 & 0.9$^{+1.2}_{-0.4}$ & 82$^{+8}_{-15}$ \\
\hline

\end{tabular}
\caption{
The best-fit S\'ersic profile parameters for nine galaxies that are relatively bright in F200W with the reported values in that filter. 
The 2nd and 3rd column show the photometric redshift and stellar mass derived from SED fitting as reported in L23. The fourth column shows the AB magnitude as derived from the aperture fluxes in L23. The 5th and 6th column are the total integrated magnitudes from \textsc{galfit} by fitting with WebbPSF and the image of a star, respectively. The 7th and 8th column are the fitted  S\'ersic indices $n$ using WebbPSF and a star cutout as the PSF, the 9th and 10th column show the corresponding effective radii $r_e$ (pc) along the major axis, and the 11th and 12th columns are the best fit values for the minor-to-major axisratio ($b/a$). 
The most right columns are the adopted $n$ and effective radii along the major axis $r_e$.
We randomly sample over the \textsc{galfit} errors and each model is then placed in the residuals of the other galaxies (excluding 16624) and fitted with two PSFs, resulting in 14 new measurements per galaxy. The values are the medians of all 16 fitted parameters with (16,84)\% quantiles as the upper and lower bound errors.
}
\label{tab:sizes-n}
\end{center}
}
\end{table*}

\section{Profile Fitting}
\subsection{Methodology}
\label{sec:methodology}
The candidate massive galaxies 
reported in L23 are fit with a  \citet{Sersic1968} profile,
    \begin{equation}
    \label{eq:radialprofile}
    I(r) = I_{\rm e} \exp{\left(-b_n\left[(r/r_{\rm e})^{1/n}-1\right]\right)},
    \end{equation}
were $I(r)$ is the surface brightness at a distance $r$ from the center, $r_{\rm e}$ is the effective (half-light) radius,
and $I_{\rm e}$ is the corresponding effective surface brightness. 

The fits are performed with \textsc{galfit} \citep{Peng2002, Peng2010x}. 
The free parameters in the fit are the ($x$, $y$) position of the source, the total integrated magnitude, the effective radius (along the major axis) ($r_{\rm e}$), the S\'ersic index ($n$), the
projected minor-to-major axis ratio ($b/a$), and the position angle. The major axis radii are circularized using $r_{\rm e,c} = r_{\rm e} \sqrt{b/a}$. 
We allow $n$ to vary between 0.5 and 20, $r_{\rm e}$ between 0.5 and 100 pixels and the total integrated magnitude between -5 and +5 difference from the aperture magnitude reported by L23.  
Each fit is performed twice, first with
the synthetic PSF and then using the empirical PSF (see Sec.\ \ref{sec:data}).

It is not immediately obvious which JWST band to use when measuring morphologies of these early red galaxies. Ideally, we would fit the galaxies in all bands to test whether there are size trends between the different bands. 

However, at the shortest wavelengths (F115W and F150W) these red galaxies are often faint, 
whereas in the long wavelength bands (F277W, F356W, and F444W) the resolution is relatively poor.
As a compromise we use the F200W band, sampling the rest-frame near-UV, for our analysis.

It is the reddest band that maintains the spatial sampling of the short wavelength camera. Four galaxies have a S/N ratio $<$5 even in F200W (L23-2859, L23-13050, L23-35300, and L23-39575). We do not include these galaxies in the analysis.\footnote{We verified that their sizes in redder bands are not obviously larger than those of the nine remaining galaxies.}

Contaminating sources are masked in the following way.
We estimate the background using sigma-clipped statistics with a filter size of 5 pixels and the background RMS.  
The background is then subtracted from the data and then convolved with a 2D Gaussian kernel with a FWHM of 3 pixels. Using this convolved background-subtracted image, we detect sources with a 1.5$\sigma$ detection threshold, where $\sigma$ is the background RMS before convolution. 
The final result is a mask map, with non-zero valued pixels that are ignored in \textsc{galfit} during the fitting procedure. 

L23-16624 has a complicated structure. In the shorter wavelength bands, we clearly observe three components. In the F444W band, these blend together into a bigger source with no substructure. This means that the central component contains most of the stellar mass. Also, the aperture diameter of 0.32'' in L23 only fully covers the central component, which is why we choose to only fit the central component and mask out the other two, using a mask threshold of 0.7$\sigma$. Yet, we caution that the photometry and derived stellar mass could be off because of this. 
We also test fitting with three S\'ersic components and this gave similar outcomes for the best fit S\'ersic parameters for the central component. 

For L23-38094, we also clearly observe two components in F115W and F150W, which also blend together as one source in the longer wavelength bands. Therefore, we fit this galaxy with two S\'ersic components and report the results of the central one.

\subsection{S\'ersic indices, sizes and uncertainties}
In Fig.\ \ref{fig:setup_segmentation} we show the individual images for the nine galaxies that are sufficiently bright in F200W. We also show the mask (as described in the previous section), the best-fit model for the WebbPSF run, and the corresponding residual for each source. 
For these nine galaxies the fits in F200W are generally excellent, with no obvious systematic residuals. 
The resulting best fit parameters, for both the synthetic and the empirical PSF,
are listed in Table \ref{tab:sizes-n}.

The best-fit circularized effective radii differ by less than $\sim15$\% when we fit with the empirical vs theoretical PSF. 
For some galaxies, the different PSFs affect the measurement of $n$, for example for L23-25666 where $n_{\mathrm{webb}}$=4.1 and  $n_{\mathrm{star}}$=1.7. Interestingly, the sizes are not very different (80\,pc and 78\,pc).

Estimating uncertainties in the S\'ersic profile parameters of faint galaxies is challenging, primarily because systematic effects typically dominate them. In this work, we measure fiducial sizes and S\'ersic indices, as
well as uncertainties on these parameters, for the nine galaxies fit in F200W in the following way. 
First, we randomly sample over the \textsc{galfit} errors and each sampled best-fit model is then placed in the residuals of each of the other galaxies (excluding 16624 because the residual has two blobs). Then we fit the galaxies again, as described above, using both the WebbPSF and the empirical PSF. 
In this way, 
each galaxy is fit in 7 different residuals and with two different PSFs, leading to $2\times 7 + 2 = 16$ different fitted models. With this approach, we include systematic uncertainties in the uncertainty budget.
The median values are then adopted as the fiducial $r_{\rm e}$ and $n$, and the
upper and lower uncertainties are given by the (16\%,84\%) quantiles, which we show in Table \ref{tab:sizes-n}.

We only allow $n$ to vary between 0.5 and 20, so if the modelling finds 0.5 multiple times, the lower uncertainty on $n$ will be 0. This happened for L23-21834 and L23-7274. In some cases (especially L23-14924) the errors can be quite large, indicating that 
some of the galaxies are barely resolved.

In the following, we assume that the observed light is galaxy-dominated and interpret the results accordingly. However, based on the morphology alone we cannot rule out that the light of some galaxies has a significant AGN contribution.

\begin{figure*}[htbp]
    \begin{center}  
    \includegraphics[trim = {0.4cm 0.7cm 0cm 0.4cm}, clip, width=0.64\textwidth]{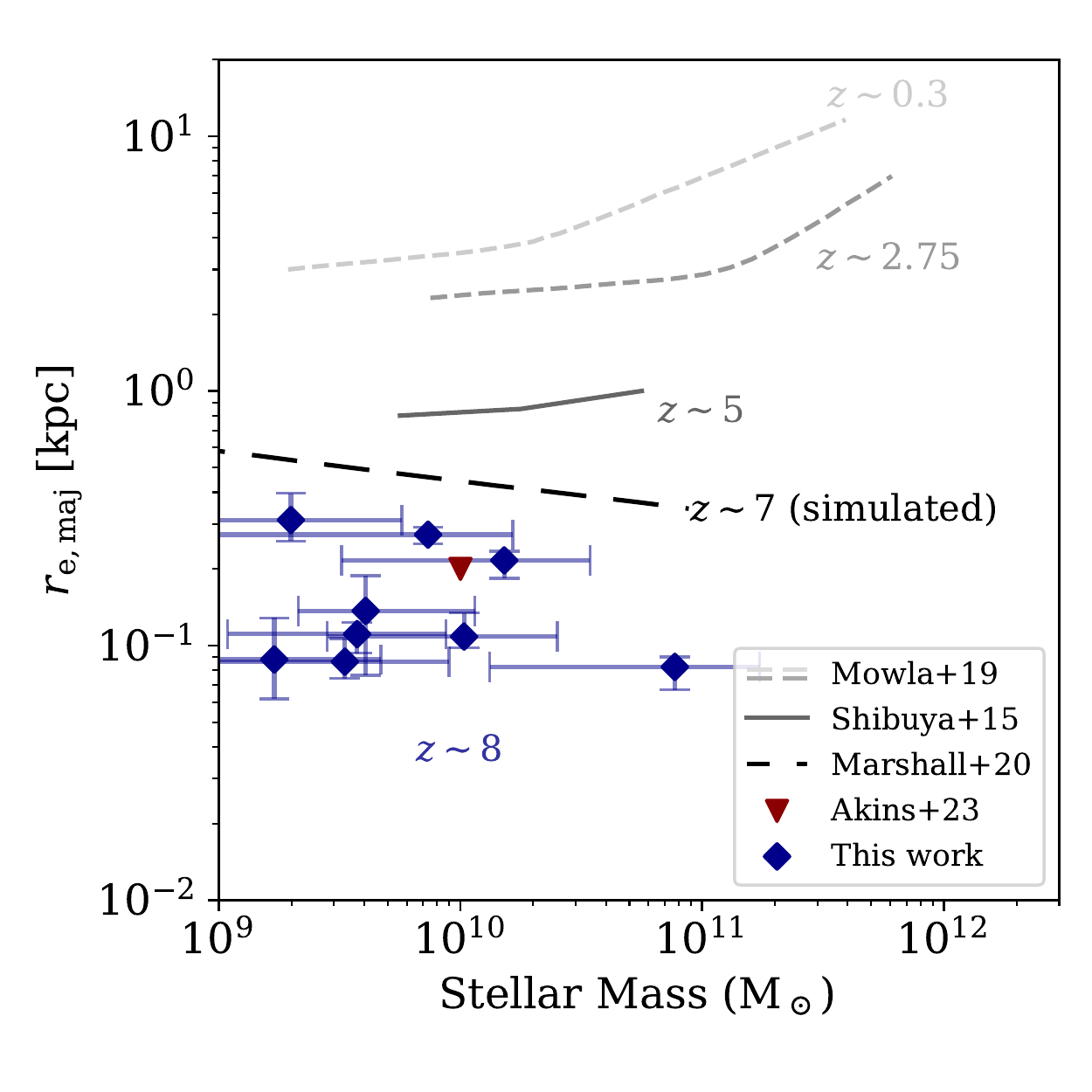}
    \end{center}
 \caption{Effective radius along major axis versus stellar mass for the 9  galaxies in F200W. The stellar masses and their errors are taken from L23.
We show the tracks for the full population of galaxies for $z\sim0.35$ (light grey), $z\sim2.75$ (grey) obtained from \citet{Mowla2019}, a size-mass relation at $z\sim5$ derived from \citet{Shibuya2015} (solid dark grey; see text) and  two massive galaxies at  $z>7$ reported by \citet{Akins2023} (red triangle; see text).
The size-mass relation of simulated galaxies at $z=7$ in the BLUETIDES cosmological hydrodynamical simulation is shown by the black dashed line \citep{Marshall2022} (see Sec.\ \ref{sec:discussion}).
}
\label{fig:sizemass}
\end{figure*}

\section{Results}
\label{sec:results}
\subsection{The Size-Mass Relation}
The main result of this study is that the
measured sizes in rest-frame UV of the nine galaxies in the structural sample are extremely small, ranging from $\sim 80$\,pc
to $\sim 300$\,pc. 
In Fig.\ \ref{fig:sizemass} we show the relation between effective radius along the major axis and stellar mass for these nine sources.  
 We also show the (rest-frame optical) size -- mass relations since $z\sim3$ derived by \citet{Mowla2019}. The light grey dashed line shows their
 broken power-law fits for galaxies $0.1<z<0.5$, which we note as $z\sim0.3$, and  for $2.5<z<3$, noted as $z\sim$ 2.75. 
In addition, we add a size-mass relation determined from \citet{Shibuya2015} at $z=5$ (Fig.\ 5 therein). For the three stellar mass bins $\log$($M_*$/M$_\odot$) = [9.5-10, 10-10.5, 10.5-11], we take the median effective radius ($r_e^{\mathrm{UV}}$) at $z=5$. 
We then plot two linear lines through to these three scatter points, as shown by the dark grey solid line. The size-mass relation of simulated galaxies at $z=7$ in the BLUETIDES cosmological hydrodynamical simulation is shown by the black long dashed line \citep{Marshall2022} (see Sec.\ \ref{sec:discussion}).

For their apparent mass and redshift, the galaxies in our structural sample are smaller
than any other galaxy population that have been observed at $0<z<5$. 
Taken at face value, our results continue a trend that has been a key result from HST over the past two decades: at fixed stellar mass,
galaxies are progressively smaller with increasing redshift \citep[e.g.,][]{Bouwens2004, Daddi2005,Trujilo2006,Trujillo2007, Buitrago2008,Dokkum2008, vanderWel2008, Williams2014}. Specifically,
the objects in our sample are about $10\times-20\times$ smaller than nearby
galaxies with the same mass. Furthermore, even if the galaxy stellar masses are overestimated by an order of magnitude, the observed sizes are smaller than typical star-forming galaxies at $0<z<5$. 
Converting the apparent magnitudes into absolute magnitudes and comparing this to the $\mathrm{M}_{\mathrm{UV}}$ - size relation at $z\sim8$ for Lyman-break galaxies \citep[e.g.][]{Yang2022}, we find that they are $\sim 3\times$ smaller than the average relation at fixed absolute magnitude.

There is some recent corroboration of
our results: we show
two massive compact galaxies at $z>7$ that were found by \citet{Akins2023}
(red triangle in Fig.\ \ref{fig:sizemass}). They find stellar masses of $M_* \sim10^{10}$M$_\odot$ and derive an upper limit for the effective radius of  $R_{\mathrm{e}}<200$ pc in F444W for both galaxies. 
These findings are consistent with the sizes and masses found in this work.

\begin{figure*}[htbp]
    \includegraphics[trim={1cm 0 2cm 0},clip, width=0.49\textwidth]{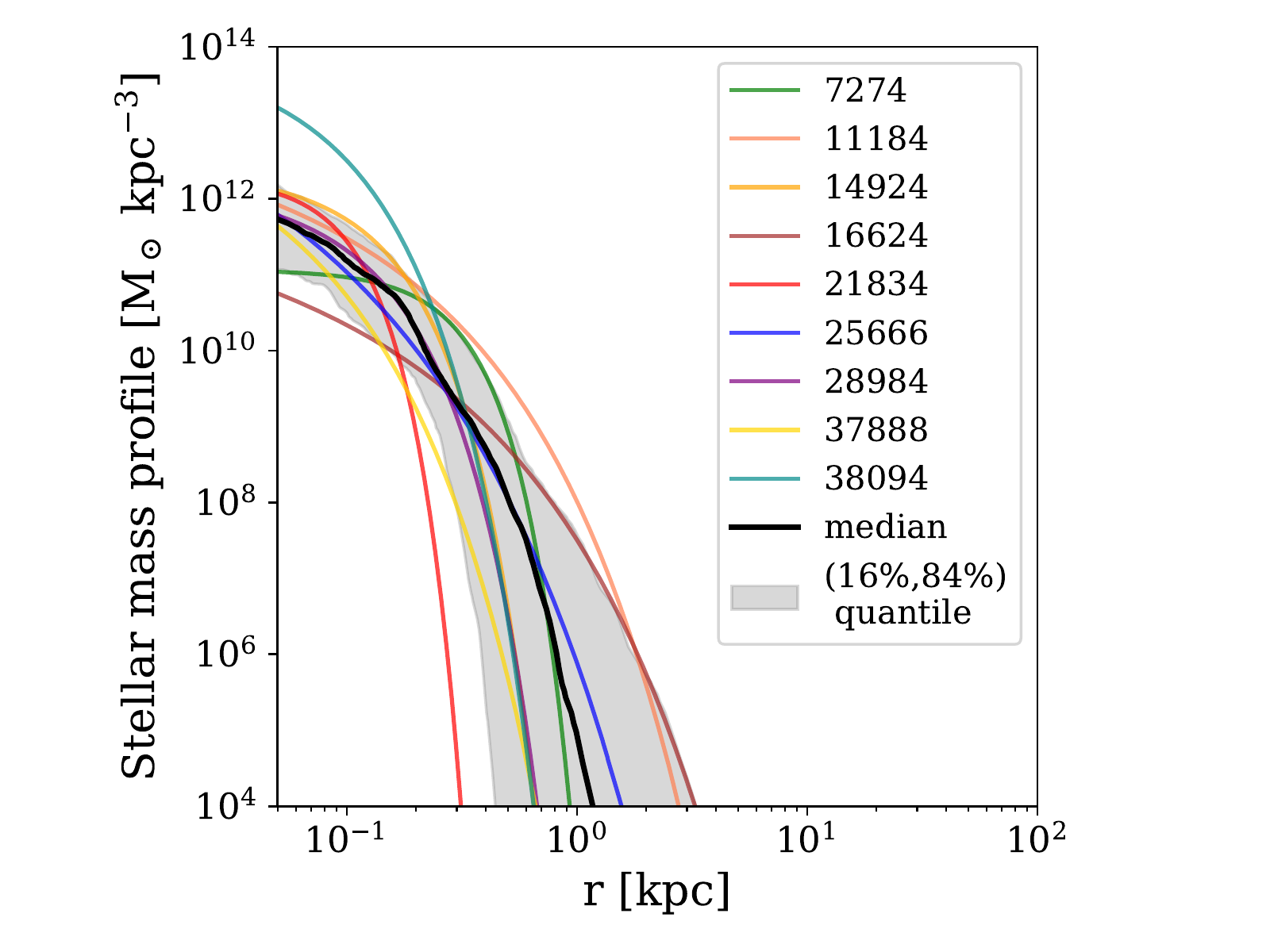}
\hfill
    \includegraphics[trim={1cm 0 2cm 0},clip, width=0.49\textwidth]{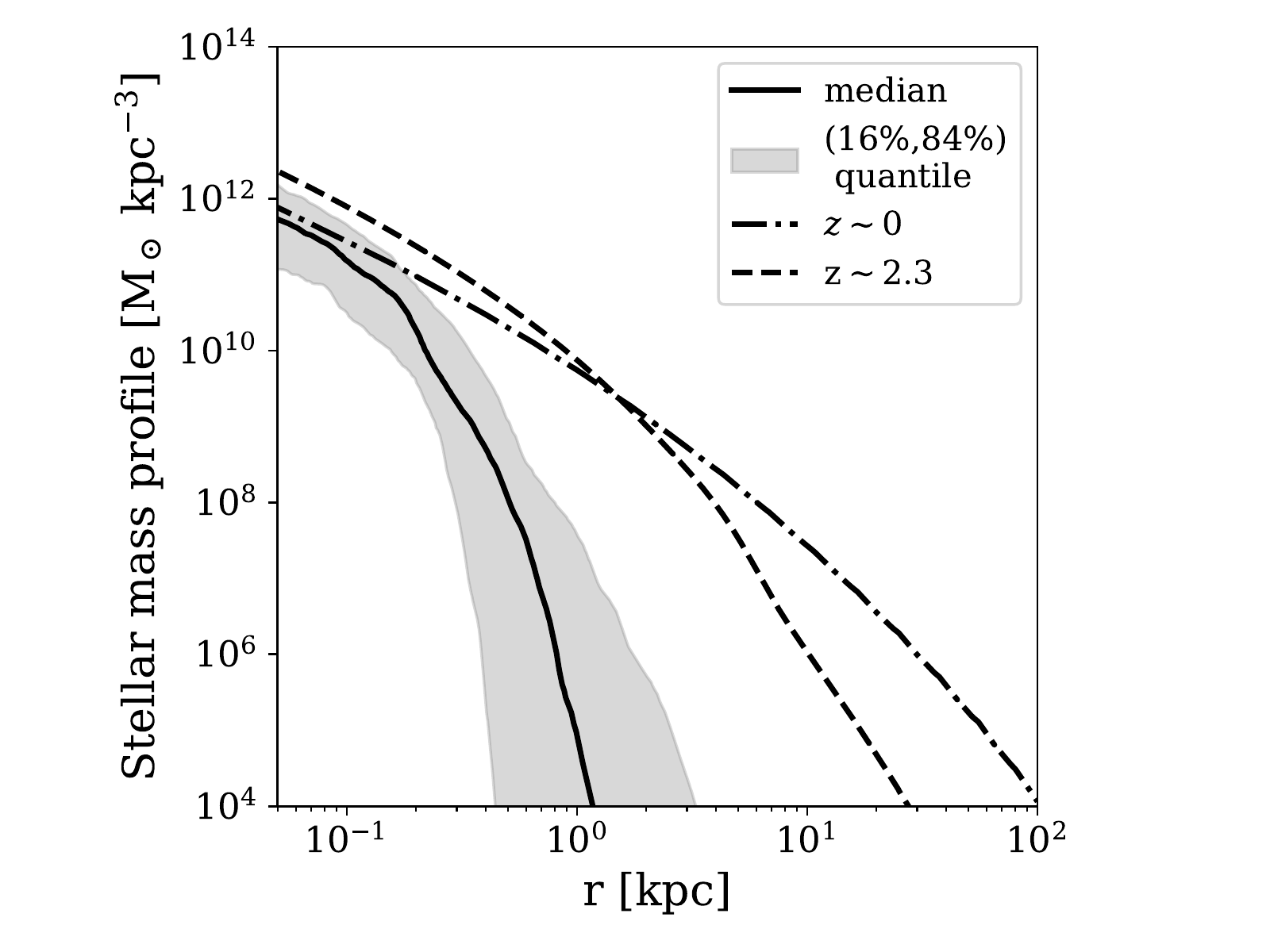}
 \caption{Left: The stellar mass profiles for the nine well-fit
 galaxies in F200W derived from $r_e$ and $n$ in Table \ref{tab:sizes-n}. The median
 profile is shown in black, along with the (16\%, 84\%) quantile range (see text).
Right: The median stellar mass profiles compared to stellar mass profiles of compact massive quiescent galaxies at $z\sim2.3$ and massive elliptical galaxies at $z=0$ obtained from \citet{Bezanson2009}. The central densities are remarkably similar, consistent with inside-out growth over the entire redshift range.}
\label{fig:massprofiles}
\end{figure*}

\subsection{Stellar Mass Profiles}
The small sizes, in combination with the assumed high stellar masses from L23, are remarkable because they imply very high stellar densities. We derive the three-dimensional stellar mass profiles for each galaxy in the structural sample, 
by performing an Abel transform to the 2D S\'ersic profile, following e.g.\ \citet{Bezanson2009}: 

\begin{equation}
\rho\left(r\right) = \frac{b_n}{\pi} \frac{I_{\rm e}}{r_{\rm e}} \left( \frac{r}{r_{\rm e}} \right)^{\left(\frac{1}{n} -1\right)} \int_1^\infty \frac{\exp\left(-b_n \left( r/r_{\rm e} \right)^{\frac{1}{n}} t\right)}{\sqrt{t^{2n} -1}}dt,
\end{equation}
with $\rho\left(r\right)$  in units of L$_\odot$\,kpc$^{-3}$.
To convert each luminosity profile into a stellar mass profile, we make a simplistic assumption that the $M/L$ ratio does not change with radius, such that the shape of the profile remains the same. The profile is then scaled such that its integral equals the total stellar mass, for which we use
\begin{equation}
    M_{*, \mathrm{fit}} = M_{*, \mathrm{L23}} \frac{L_{\mathrm{fit}}}{L_{\mathrm{L23}}},
    \end{equation}
that is, we correct the masses in L23 so that the sizes and S\'ersic indices from \textsc{galfit} are self-consistent with the total luminosities and masses.

The stellar mass profiles for the nine galaxies are shown in Fig.\ \ref{fig:massprofiles} (left).
From the $16\times 9$ stellar mass profiles we take the median (shown in black) and (16\%, 84\%) quantile range (shown in grey).  As expected, the stellar mass
densities are extremely high, reaching $\sim 10^{12}$\,M$_{\odot}$\,kpc$^{-3}$
in the central regions. The density falls of very rapidly with radius, dropping below $\sim 10^8$\,M$_{\odot}$\,kpc$^{-3}$
at $r\sim 500$\,pc.  In the following section we compare the profiles to those of plausible descendants at lower redshifts.

\subsection{Comparison to Stellar Mass Profiles of Massive Galaxies at Later Times}

An important question is whether the galaxies are {\em too} compact, that is,
whether the stellar densities exceed those of all plausible descendants.
We can expect that the most massive galaxies at high redshift evolve into the most massive galaxies at lower redshifts, and that progenitors and descendants can be approximately matched by their cumulative number density
\citep[e.g.,][]{Dokkum2010, Papovich2011, Leja2013,Jaacks2016, Torrey2017}. 
In Fig.\ \ref{fig:massprofiles} (right) we compare\footnote{assuming that the measured size in rest-frame UV is equal\\ to rest-frame optical} the median stellar mass profile at $z\sim 8$ to average stellar mass density profiles of 
nine massive ($M_* \geq 1\times10^{11}$M$_\odot$) quiescent galaxies at $z=2.3$ studied by \citet{Kriek2006}, \citet{Dokkum2008} and \citet{Bezanson2009} (dashed line) and to
$z\approx0$ elliptical galaxies ($M_* > 5\times10^{11}$M$_\odot$) from \citet{Tal2009} (dash-dotted line). 
All three galaxy populations have
a cumulative number density of 
a few $\times 10^{-5}$\,Mpc$^{-3}$. 
As expected, there is a large difference between the stellar mass densities at larger radii (r $>$1 kpc), where nearby ellipticals and $z=2.3$ compact galaxies have densities that are at least an order of magnitude higher.
However, the median stellar mass densities in the innermost regions (at $r\sim$100 pc) are very similar 
for all populations, and we infer that the extreme stellar densities
that we find are similar to those of the cores of the most massive systems
throughout cosmic time.

\section{Discussion}
\label{sec:discussion}
The central result of our paper is that the  candidate massive galaxies at $z\sim 8$ identified in L23
are extremely compact, at least at rest-frame
wavelengths of $\sim 0.25\,\mu$m. 
We find that they have sizes of $\sim 150$\,pc, making them
smaller than $z=0$ galaxies of the same mass by at least an order of magnitude.

Yet, there are a few caveats that need to be considered. Firstly, there may be systematic biases in the \textsc{galfit} sizes. \citet{Ono2022} inject models of various sizes for a range of magnitudes into the noise map and examine the differences between the input and output sizes. They find that the output sizes are underestimated among the faint sources, but that the effect is small  for sources with magnitudes $\sim$27-28 in F200W. We also assessed the uncertainties by placing models with a wide range of radii and magnitudes in empty areas and recovering their sizes. This method is well-suited for determining systematic differences between true and measured sizes, but unlike our default method (which uses the actual residuals) it does not take differences between real galaxies and S\'ersic profiles into account. We find that there is a systematic bias for fainter and larger objects, with measured sizes often underestimating the true sizes, consistent with \citet{Ono2022}. However, for the majority of sources in our data (7/9 sources with mag $<$ 28 and pixel sizes $<$ 4), the effect is small (11\% scatter). For the two fainter objects (mag $>$ 28 and pixel sizes $<2$), the systematic scatter is larger (24\%), but this scatter is still much smaller than the errors measured with the default method ($>$43\%).
Another important caveat is that the size measurement may depend on wavelength. Initial studies suggest that the effect is small: \citet{Yang2022} find a ratio of
 ($r_{\mathrm{e,F444W}}/r_{\mathrm{e,F200W}} = 1.15$) for $z\sim$7-9 galaxies. However, the effect could be larger for these massive galaxies.

Although it seems difficult to measure sizes this small, we note
that it is equivalent to measuring sizes of $\sim 600$\,pc with Hubble
at $z\sim 2$ when the differences in
angular diameter distance and resolving
power are taken into account. Sizes down to $\sim 400$\,pc have
been measured routinely with HST \citep[see, e.g.,][]{Szomoru2010}. This is also in line with recent results from \citet{Yang2022}, who showed that NIRCam imaging data are of sufficient quality to measure effective radii down to $\sim$ 100 pc at z $>$ 7.

The stellar densities are correspondingly high, reaching $\sim 10^{12}$\,M$_{\odot}$\,kpc$^{-3}$ in the central regions.
While extreme, as shown in Fig.\ \ref{fig:massprofiles},
these densities are similar to those of their plausible descendants.
The most straightforward interpretation is that 
the central $\sim 100$\,pc of massive galaxies formed very early, at $z\sim 8$, and that the subsequent growth
was largely inside-out, through star formation in disks or through minor mergers.
This extends the arguments made for $z=2-3$ galaxies in \citet{Bezanson2009, Naab2009,Hopkins2010,trujillo2011,Barro2013,Dokkum2015} and
\citet{DelaRosa2016} all the way to $z\sim 8$ and the central $\sim 100$\,pc. This result is consistent with the abundance matching constraints by \citet{Moster2020} who found that $M_* \sim 10^{9.5}$M$_\odot$ galaxies at $z\sim8$ are progenitors of present day  $ M_* >10^{11}$M$_\odot$ galaxies for number densities in the range of $10^{-6} - 10^{-5}$Mpc$^{-3}$.

Turning to the assembly of the $z\sim 8$ galaxies, 
the small sizes and high densities
are qualitatively consistent with extremely efficient dissipative collapse, as required by
the stellar masses and number density estimates of the galaxies \citep{BoylanKolchin2023}.
As the galaxies are already at the upper end of what can exist at these early times, their star
formation rates will likely decrease soon. This is also required by the density profiles in Fig.\ \ref{fig:massprofiles}:
if the central densities grow further by more than a factor of $\sim 2$ they would begin to exceed those of
their likely low redshift descendants.

If the galaxies quench shortly after $z\sim 8$ it would produce a population of early quiescent galaxies. Such
galaxies may have already been observed; specifically, \citet{Carnall2023} find massive quiescent galaxies at $3 < z < 5$ and argue that some of these likely had
masses of $\log$($M_*$/M$_\odot$$)>10$ at $z\geq 8$.

The early formation of the central 
$\sim 100$\,pc suggests that supermassive black holes in massive early-type
galaxies also formed very early. In fact, the formation of the black hole and the stars may not have been
entirely separable events.
We calculate the sphere of influence for the median stellar mass profile with $M_{\star}(r<r_{h})=2M_{\rm{BH}}$, where we use black hole masses of $M_{\rm{BH}} = 10^8$M$_\odot$ and $10^9$\,M$_{\odot}$.  This gives $r_{\mathrm{h}}= 28$\,pc
for $M_{\rm BH}=10^8$\,M$_{\odot}$ and 114\,pc for $M_{\rm BH}=10^9$\,M$_{\odot}$. 
These values are in the same range as the half-light radii of the galaxies.

It will be interesting to see if the strong size evolution that is implied by our results is reproduced in simulations.
The most obvious comparison is with \citet{Marshall2022}, who made predictions for sizes of galaxies observed with JWST at $z=7-11$ using the BLUETIDES cosmological hydrodynamical simulation. 
In Fig.\ \ref{fig:sizemass} we show the half mass radius - stellar mass relation that they predict for $z=7$ for a mass-limited sample. 
Our results fall below their predictions.

We also briefly explore the \textsc{Hydrangea} simulations \citep[see a detailed description in][]{Bahe_2017}, which is part of the \textsc{c-eagle} simulations, a set of zoom-in galaxy clusters. Galaxies with similar half mass radii and stellar masses are present in these simulations at $z\approx7$, where they are the cores of future massive ellipticals in clusters. This means that we might be observing the highest peaks of the underlying dark matter distribution. 
Larger area surveys with JWST, combined with spectroscopic redshifts, will show whether we happen to observe
a large overdensity at $z\sim 8$ \citep[see for example][]{Hashimoto2023}.

The redshifts and stellar masses of the galaxies remain the main sources
of uncertainty. As noted in L23, the galaxies are likely a mix of
$z\sim 5-7$ AGN, objects dominated by extreme and exotic emission
lines, and genuine massive galaxies at $z\sim 8$. 

From the morphologies alone, we cannot distinguish between these scenarios \citep[see e.g.][who show that the measured morphology depends on the contrast between the AGN and the host galaxy]{Harikane2023}. 
Examining our fits, 
the most compact sources, L23-25666 and L23-37888 and L23-38094 could be dominated by
a point source, and the extreme apparent density of L23-38094 suggests that
this object may either be contaminated by an AGN or an extreme emission line object.   
JWST spectroscopy is needed to make progress on these issues.

\begin{acknowledgements}
We are grateful for the observations made with the NASA/ESA/CSA James Webb Space Telescope and the CEERS (PID: 1345) team. The CEERS data are publicly available in the Mikulski Archive for Space Telescopes (MAST) archive at the Space Telescope Science Institute 
(DOI: \dataset[https://doi.org/10.17909/z7p0-8481]{https://doi.org/10.17909/z7p0-8481}). STScI is operated by the Association of Universities for Research in Astronomy, Inc., under NASA contract NAS5–26555.  The \textsc{grizli} pipeline was used to reduce the data, and the resulting data products can be accessed by anyone through the Dawn JWST Archive (DJA). DJA is an initiative of the Cosmic Dawn Center, which is funded by the Danish National Research Foundation under grant No.\ 140.
Finally, we express our gratitude to the anonymous referee for their useful comments on this paper, which significantly enhanced the quality of our work. 
\end{acknowledgements}

\newpage
\bibliography{paper}{}

\begin{thebibliography}{}
\expandafter\ifx\csname natexlab\endcsname\relax\def\natexlab#1{#1}\fi
\providecommand{\url}[1]{\href{#1}{#1}}
\providecommand{\dodoi}[1]{doi:~\href{http://doi.org/#1}{\nolinkurl{#1}}}
\providecommand{\doeprint}[1]{\href{http://ascl.net/#1}{\nolinkurl{http://ascl.net/#1}}}
\providecommand{\doarXiv}[1]{\href{https://arxiv.org/abs/#1}{\nolinkurl{https://arxiv.org/abs/#1}}}

\bibitem[{{Akins} {et~al.}(2023){Akins}, {Casey}, {Allen}, {Bagley},
  {Dickinson}, {Finkelstein}, {Franco}, {Harish}, {Arrabal Haro}, {Ilbert},
  {Kartaltepe}, {Koekemoer}, {Liu}, {Long}, {McCracken}, {Paquereau},
  {Papovich}, {Pirzkal}, {Rhodes}, {Robertson}, {Shuntov}, {Toft}, {Yang},
  {Barro}, {Bisigello}, {Buat}, {Champagne}, {Cooper}, {Costantin}, {de la
  Vega}, {Drakos}, {Faisst}, {Fontana}, {Fujimoto}, {Gillman},
  {G{\'o}mez-Guijarro}, {Gozaliasl}, {Hathi}, {Hayward}, {Hirschmann},
  {Holwerda}, {Jin}, {Kocevski}, {Kokorev}, {Lambrides}, {Lucas}, {Magdis},
  {Magnelli}, {McKinney}, {Mobasher}, {P{\'e}rez-Gonz{\'a}lez}, {Rich},
  {Seill{\'e}}, {Talia}, {Urry}, {Valentino}, {Whitaker}, {Yung}, \&
  {Zavala}}]{Akins2023}
{Akins}, H.~B., {Casey}, C.~M., {Allen}, N., {et~al.} 2023, arXiv e-prints,
  arXiv:2304.12347, \dodoi{10.48550/arXiv.2304.12347}

\bibitem[{{Atek} {et~al.}(2023){Atek}, {Shuntov}, {Furtak}, {Richard}, {Kneib},
  {Mahler}, {Zitrin}, {McCracken}, {Charlot}, {Chevallard}, \&
  {Chemerynska}}]{Atek2023}
{Atek}, H., {Shuntov}, M., {Furtak}, L.~J., {et~al.} 2023, \mnras, 519, 1201,
  \dodoi{10.1093/mnras/stac3144}

\bibitem[{{Austin} {et~al.}(2023){Austin}, {Adams}, {Conselice}, {Harvey},
  {Ormerod}, {Trussler}, {Li}, {Ferreira}, {Dayal}, \&
  {Juod{\v{z}}balis}}]{Austin2023}
{Austin}, D., {Adams}, N., {Conselice}, C.~J., {et~al.} 2023, \apjl, 952, L7,
  \dodoi{10.3847/2041-8213/ace18d}

\bibitem[{{Bah{\'e}} {et~al.}(2017){Bah{\'e}}, {Barnes}, {Dalla Vecchia},
  {Kay}, {White}, {McCarthy}, {Schaye}, {Bower}, {Crain}, {Theuns}, {Jenkins},
  {McGee}, {Schaller}, {Thomas}, \& {Trayford}}]{Bahe_2017}
{Bah{\'e}}, Y.~M., {Barnes}, D.~J., {Dalla Vecchia}, C., {et~al.} 2017, \mnras,
  470, 4186, \dodoi{10.1093/mnras/stx1403}

\bibitem[{{Barro} {et~al.}(2013){Barro}, {Faber}, {P{\'e}rez-Gonz{\'a}lez},
  {Koo}, {Williams}, {Kocevski}, {Trump}, {Mozena}, {McGrath}, {van der Wel},
  {Wuyts}, {Bell}, {Croton}, {Ceverino}, {Dekel}, {Ashby}, {Cheung},
  {Ferguson}, {Fontana}, {Fang}, {Giavalisco}, {Grogin}, {Guo}, {Hathi},
  {Hopkins}, {Huang}, {Koekemoer}, {Kartaltepe}, {Lee}, {Newman}, {Porter},
  {Primack}, {Ryan}, {Rosario}, {Somerville}, {Salvato}, \& {Hsu}}]{Barro2013}
{Barro}, G., {Faber}, S.~M., {P{\'e}rez-Gonz{\'a}lez}, P.~G., {et~al.} 2013,
  \apj, 765, 104, \dodoi{10.1088/0004-637X/765/2/104}

\bibitem[{{Barro} {et~al.}(2023){Barro}, {Perez-Gonzalez}, {Kocevski},
  {McGrath}, {Trump}, {Simons}, {Somerville}, {Yung}, {Arrabal Haro}, {Bagley},
  {Cleri}, {Costantin}, {Davis}, {Dickinson}, {Finkelstein}, {Giavalisco},
  {Gomez-Guijarro}, {Hathi}, {Hirschmann}, {Akins}, {Holwerda},
  {Huertas-Company}, {Lucas}, {Papovich}, {Seille}, {Tacchella}, {Wilkins}, {de
  la Vega}, {Yang}, \& {Zavala}}]{Barro2023}
{Barro}, G., {Perez-Gonzalez}, P.~G., {Kocevski}, D.~D., {et~al.} 2023, arXiv
  e-prints, arXiv:2305.14418, \dodoi{10.48550/arXiv.2305.14418}

\bibitem[{{Bezanson} {et~al.}(2009){Bezanson}, {van Dokkum}, {Tal},
  {Marchesini}, {Kriek}, {Franx}, \& {Coppi}}]{Bezanson2009}
{Bezanson}, R., {van Dokkum}, P.~G., {Tal}, T., {et~al.} 2009, \apj, 697, 1290,
  \dodoi{10.1088/0004-637X/697/2/1290}

\bibitem[{{Bouwens} {et~al.}(2004){Bouwens}, {Illingworth}, {Blakeslee},
  {Broadhurst}, \& {Franx}}]{Bouwens2004}
{Bouwens}, R.~J., {Illingworth}, G.~D., {Blakeslee}, J.~P., {Broadhurst},
  T.~J., \& {Franx}, M. 2004, \apjl, 611, L1, \dodoi{10.1086/423786}

\bibitem[{{Boyett} {et~al.}(2023){Boyett}, {Trenti}, {Leethochawalit},
  {Calabr{\'o}}, {Metha}, {Roberts-Borsani}, {Dalmasso}, {Yang}, {Santini},
  {Treu}, {Jones}, {Henry}, {Mason}, {Morishita}, {Nanayakkara}, {Roy}, {Wang},
  {Fontana}, {Merlin}, {Castellano}, {Paris}, {Bradac}, {Marchesini}, {Mascia},
  {Pentericci}, {Vanzella}, \& {Vulcani}}]{Boyett2023}
{Boyett}, K., {Trenti}, M., {Leethochawalit}, N., {et~al.} 2023, arXiv
  e-prints, arXiv:2303.00306, \dodoi{10.48550/arXiv.2303.00306}

\bibitem[{{Boylan-Kolchin}(2023)}]{BoylanKolchin2023}
{Boylan-Kolchin}, M. 2023, Nature Astronomy, \dodoi{10.1038/s41550-023-01937-7}

\bibitem[{{Brammer}(2023)}]{gabe_brammer_grizli}
{Brammer}, G. 2023, grizli, 1.5.2,  Zenodo, \dodoi{10.5281/ZENODO.1146904}

\bibitem[{Brammer(2023)}]{mosaicsV4}
Brammer, G. 2023, JWST image mosaics grizli-v4,  University of Copenhagen,
  \dodoi{10.17894/UCPH.E3D897AF-233A-4F01-A893-7B0FAD1F66C2}

\bibitem[{{Brammer} {et~al.}(2008){Brammer}, {van Dokkum}, \&
  {Coppi}}]{Brammer2008}
{Brammer}, G.~B., {van Dokkum}, P.~G., \& {Coppi}, P. 2008, \apj, 686, 1503,
  \dodoi{10.1086/591786}

\bibitem[{{Buitrago} {et~al.}(2008){Buitrago}, {Trujillo}, {Conselice},
  {Bouwens}, {Dickinson}, \& {Yan}}]{Buitrago2008}
{Buitrago}, F., {Trujillo}, I., {Conselice}, C.~J., {et~al.} 2008, \apjl, 687,
  L61, \dodoi{10.1086/592836}

\bibitem[{{Carnall} {et~al.}(2018){Carnall}, {McLure}, {Dunlop}, \&
  {Dav{\'e}}}]{Carnall2018}
{Carnall}, A.~C., {McLure}, R.~J., {Dunlop}, J.~S., \& {Dav{\'e}}, R. 2018,
  \mnras, 480, 4379, \dodoi{10.1093/mnras/sty2169}

\bibitem[{{Carnall} {et~al.}(2023){Carnall}, {McLeod}, {McLure}, {Dunlop},
  {Begley}, {Cullen}, {Donnan}, {Hamadouche}, {Jewell}, {Jones}, {Pollock}, \&
  {Wild}}]{Carnall2023}
{Carnall}, A.~C., {McLeod}, D.~J., {McLure}, R.~J., {et~al.} 2023, \mnras, 520,
  3974, \dodoi{10.1093/mnras/stad369}

\bibitem[{{Castellano} {et~al.}(2022){Castellano}, {Fontana}, {Treu},
  {Santini}, {Merlin}, {Leethochawalit}, {Trenti}, {Vanzella}, {Mestric},
  {Bonchi}, {Belfiori}, {Nonino}, {Paris}, {Polenta}, {Roberts-Borsani},
  {Boyett}, {Brada{\v{c}}}, {Calabr{\`o}}, {Glazebrook}, {Grillo}, {Mascia},
  {Mason}, {Mercurio}, {Morishita}, {Nanayakkara}, {Pentericci}, {Rosati},
  {Vulcani}, {Wang}, \& {Yang}}]{Castellano2022}
{Castellano}, M., {Fontana}, A., {Treu}, T., {et~al.} 2022, \apjl, 938, L15,
  \dodoi{10.3847/2041-8213/ac94d0}

\bibitem[{{Daddi} {et~al.}(2005){Daddi}, {Renzini}, {Pirzkal}, {Cimatti},
  {Malhotra}, {Stiavelli}, {Xu}, {Pasquali}, {Rhoads}, {Brusa}, {di Serego
  Alighieri}, {Ferguson}, {Koekemoer}, {Moustakas}, {Panagia}, \&
  {Windhorst}}]{Daddi2005}
{Daddi}, E., {Renzini}, A., {Pirzkal}, N., {et~al.} 2005, \apj, 626, 680,
  \dodoi{10.1086/430104}

\bibitem[{{de la Rosa} {et~al.}(2016){de la Rosa}, {La Barbera}, {Ferreras},
  {S{\'a}nchez Almeida}, {Dalla Vecchia}, {Mart{\'\i}nez-Valpuesta}, \&
  {Stringer}}]{DelaRosa2016}
{de la Rosa}, I.~G., {La Barbera}, F., {Ferreras}, I., {et~al.} 2016, \mnras,
  457, 1916, \dodoi{10.1093/mnras/stw130}

\bibitem[{{Ding} {et~al.}(2022){Ding}, {Silverman}, \& {Onoue}}]{Ding2022}
{Ding}, X., {Silverman}, J.~D., \& {Onoue}, M. 2022, \apjl, 939, L28,
  \dodoi{10.3847/2041-8213/ac9c02}

\bibitem[{{Donnan} {et~al.}(2023){Donnan}, {McLeod}, {Dunlop}, {McLure},
  {Carnall}, {Begley}, {Cullen}, {Hamadouche}, {Bowler}, {Magee}, {McCracken},
  {Milvang-Jensen}, {Moneti}, \& {Targett}}]{Donnan2023}
{Donnan}, C.~T., {McLeod}, D.~J., {Dunlop}, J.~S., {et~al.} 2023, \mnras, 518,
  6011, \dodoi{10.1093/mnras/stac3472}

\bibitem[{{Endsley} {et~al.}(2023){Endsley}, {Stark}, {Whitler}, {Topping},
  {Chen}, {Plat}, {Chisholm}, \& {Charlot}}]{Endsley2023}
{Endsley}, R., {Stark}, D.~P., {Whitler}, L., {et~al.} 2023, \mnras,
  \dodoi{10.1093/mnras/stad1919}

\bibitem[{{Finkelstein} {et~al.}(2023{\natexlab{a}}){Finkelstein}, {Bagley}, \&
  {Yang}}]{doiCEERSmosaic}
{Finkelstein}, S.~L., {Bagley}, M.~B., \& {Yang}, G. 2023{\natexlab{a}}, Data
  from The Cosmic Evolution Early Release Science Survey (CEERS),  STScI/MAST,
  \dodoi{10.17909/Z7P0-8481}

\bibitem[{{Finkelstein} {et~al.}(2022){Finkelstein}, {Bagley}, {Haro},
  {Dickinson}, {Ferguson}, {Kartaltepe}, {Papovich}, {Burgarella}, {Kocevski},
  {Huertas-Company}, {Iyer}, {Koekemoer}, {Larson}, {P{\'e}rez-Gonz{\'a}lez},
  {Rose}, {Tacchella}, {Wilkins}, {Chworowsky}, {Medrano}, {Morales},
  {Somerville}, {Yung}, {Fontana}, {Giavalisco}, {Grazian}, {Grogin}, {Kewley},
  {Kirkpatrick}, {Kurczynski}, {Lotz}, {Pentericci}, {Pirzkal}, {Ravindranath},
  {Ryan}, {Trump}, {Yang}, {Almaini}, {Amor{\'\i}n}, {Annunziatella},
  {Backhaus}, {Barro}, {Behroozi}, {Bell}, {Bhatawdekar}, {Bisigello}, {Bromm},
  {Buat}, {Buitrago}, {Calabr{\`o}}, {Casey}, {Castellano}, {Ch{\'a}vez Ortiz},
  {Ciesla}, {Cleri}, {Cohen}, {Cole}, {Cooke}, {Cooper}, {Cooray}, {Costantin},
  {Cox}, {Croton}, {Daddi}, {Dav{\'e}}, {de La Vega}, {Dekel}, {Elbaz},
  {Estrada-Carpenter}, {Faber}, {Fern{\'a}ndez}, {Finkelstein}, {Freundlich},
  {Fujimoto}, {Garc{\'\i}a-Argum{\'a}nez}, {Gardner}, {Gawiser},
  {G{\'o}mez-Guijarro}, {Guo}, {Hamblin}, {Hamilton}, {Hathi}, {Holwerda},
  {Hirschmann}, {Hutchison}, {Jaskot}, {Jha}, {Jogee}, {Juneau}, {Jung},
  {Kassin}, {Le Bail}, {Leung}, {Lucas}, {Magnelli}, {Mantha}, {Matharu},
  {McGrath}, {McIntosh}, {Merlin}, {Mobasher}, {Newman}, {Nicholls}, {Pandya},
  {Rafelski}, {Ronayne}, {Santini}, {Seill{\'e}}, {Shah}, {Shen}, {Simons},
  {Snyder}, {Stanway}, {Straughn}, {Teplitz}, {Vanderhoof}, {Vega-Ferrero},
  {Wang}, {Weiner}, {Willmer}, {Wuyts}, {Zavala}, \& {CEERS
  Team}}]{Finkelstein2022}
{Finkelstein}, S.~L., {Bagley}, M.~B., {Haro}, P.~A., {et~al.} 2022, \apjl,
  940, L55, \dodoi{10.3847/2041-8213/ac966e}

\bibitem[{{Finkelstein} {et~al.}(2023{\natexlab{b}}){Finkelstein}, {Bagley},
  {Ferguson}, {Wilkins}, {Kartaltepe}, {Papovich}, {Yung}, {Arrabal Haro},
  {Behroozi}, {Dickinson}, {Kocevski}, {Koekemoer}, {Larson}, {Le Bail},
  {Morales}, {P{\'e}rez-Gonz{\'a}lez}, {Burgarella}, {Dav{\'e}}, {Hirschmann},
  {Somerville}, {Wuyts}, {Bromm}, {Casey}, {Fontana}, {Fujimoto}, {Gardner},
  {Giavalisco}, {Grazian}, {Grogin}, {Hathi}, {Hutchison}, {Jha}, {Jogee},
  {Kewley}, {Kirkpatrick}, {Long}, {Lotz}, {Pentericci}, {Pierel}, {Pirzkal},
  {Ravindranath}, {Ryan}, {Trump}, {Yang}, {Bhatawdekar}, {Bisigello}, {Buat},
  {Calabr{\`o}}, {Castellano}, {Cleri}, {Cooper}, {Croton}, {Daddi}, {Dekel},
  {Elbaz}, {Franco}, {Gawiser}, {Holwerda}, {Huertas-Company}, {Jaskot},
  {Leung}, {Lucas}, {Mobasher}, {Pandya}, {Tacchella}, {Weiner}, \&
  {Zavala}}]{Finkelstein2023}
{Finkelstein}, S.~L., {Bagley}, M.~B., {Ferguson}, H.~C., {et~al.}
  2023{\natexlab{b}}, \apjl, 946, L13, \dodoi{10.3847/2041-8213/acade4}

\bibitem[{{Fujimoto} {et~al.}(2023){Fujimoto}, {Arrabal Haro}, {Dickinson},
  {Finkelstein}, {Kartaltepe}, {Larson}, {Burgarella}, {Bagley}, {Behroozi},
  {Chworowsky}, {Hirschmann}, {Trump}, {Wilkins}, {Yung}, {Koekemoer},
  {Papovich}, {Pirzkal}, {Ferguson}, {Fontana}, {Grogin}, {Grazian}, {Kewley},
  {Kocevski}, {Lotz}, {Pentericci}, {Ravindranath}, {Somerville}, {Wilkins},
  {Amor{\'\i}n}, {Backhaus}, {Calabr{\`o}}, {Casey}, {Cooper}, {Fern{\'a}ndez},
  {Franco}, {Giavalisco}, {Hathi}, {Harish}, {Hutchison}, {Iyer}, {Jung},
  {Lucas}, \& {Zavala}}]{Fujimoto2023}
{Fujimoto}, S., {Arrabal Haro}, P., {Dickinson}, M., {et~al.} 2023, \apjl, 949,
  L25, \dodoi{10.3847/2041-8213/acd2d9}

\bibitem[{{Furtak} {et~al.}(2023){Furtak}, {Zitrin}, {Plat}, {Fujimoto},
  {Wang}, {Nelson}, {Labb{\'e}}, {Bezanson}, {Brammer}, {van Dokkum},
  {Endsley}, {Glazebrook}, {Greene}, {Leja}, {Price}, {Smit}, {Stark},
  {Weaver}, {Whitaker}, {Atek}, {Chevallard}, {Curtis-Lake}, {Dayal}, {Feltre},
  {Franx}, {Fudamoto}, {Marchesini}, {Mowla}, {Pan}, {Suess},
  {Vidal-Garc{\'\i}a}, \& {Williams}}]{Furtak2022a}
{Furtak}, L.~J., {Zitrin}, A., {Plat}, A., {et~al.} 2023, \apj, 952, 142,
  \dodoi{10.3847/1538-4357/acdc9d}

\bibitem[{{Gong} {et~al.}(2023){Gong}, {Yue}, {Cao}, \& {Chen}}]{Gong2023}
{Gong}, Y., {Yue}, B., {Cao}, Y., \& {Chen}, X. 2023, \apj, 947, 28,
  \dodoi{10.3847/1538-4357/acc109}

\bibitem[{{Harikane} {et~al.}(2023){Harikane}, {Zhang}, {Nakajima}, {Ouchi},
  {Isobe}, {Ono}, {Hatano}, {Xu}, \& {Umeda}}]{Harikane2023}
{Harikane}, Y., {Zhang}, Y., {Nakajima}, K., {et~al.} 2023, arXiv e-prints,
  arXiv:2303.11946, \dodoi{10.48550/arXiv.2303.11946}

\bibitem[{{Hashimoto} {et~al.}(2023){Hashimoto}, {{\'A}lvarez-M{\'a}rquez},
  {Fudamoto}, {Colina}, {Inoue}, {Nakazato}, {Ceverino}, {Yoshida},
  {Costantin}, {Sugahara}, {G{\'o}mez}, {Blanco-Prieto}, {Mawatari}, {Arribas},
  {Marques-Chaves}, {Pereira-Santaella}, {Bakx}, {Hagimoto}, {Hashigaya},
  {Matsuo}, {Tamura}, {Usui}, \& {Ren}}]{Hashimoto2023}
{Hashimoto}, T., {{\'A}lvarez-M{\'a}rquez}, J., {Fudamoto}, Y., {et~al.} 2023,
  \apjl, 955, L2, \dodoi{10.3847/2041-8213/acf57c}

\bibitem[{{Haslbauer} {et~al.}(2022){Haslbauer}, {Kroupa}, {Zonoozi}, \&
  {Haghi}}]{Haslbauer2022}
{Haslbauer}, M., {Kroupa}, P., {Zonoozi}, A.~H., \& {Haghi}, H. 2022, \apjl,
  939, L31, \dodoi{10.3847/2041-8213/ac9a50}

\bibitem[{{Hopkins} {et~al.}(2010){Hopkins}, {Bundy}, {Hernquist}, {Wuyts}, \&
  {Cox}}]{Hopkins2010}
{Hopkins}, P.~F., {Bundy}, K., {Hernquist}, L., {Wuyts}, S., \& {Cox}, T.~J.
  2010, \mnras, 401, 1099, \dodoi{10.1111/j.1365-2966.2009.15699.x}

\bibitem[{{H{\"u}tsi} {et~al.}(2023){H{\"u}tsi}, {Raidal}, {Urrutia},
  {Vaskonen}, \& {Veerm{\"a}e}}]{Hutsi2023}
{H{\"u}tsi}, G., {Raidal}, M., {Urrutia}, J., {Vaskonen}, V., \& {Veerm{\"a}e},
  H. 2023, \prd, 107, 043502, \dodoi{10.1103/PhysRevD.107.043502}

\bibitem[{{Jaacks} {et~al.}(2016){Jaacks}, {Finkelstein}, \&
  {Nagamine}}]{Jaacks2016}
{Jaacks}, J., {Finkelstein}, S.~L., \& {Nagamine}, K. 2016, \apj, 817, 174,
  \dodoi{10.3847/0004-637X/817/2/174}

\bibitem[{{Jiao} {et~al.}(2023){Jiao}, {Brandenberger}, \&
  {Refregier}}]{Jiao2023}
{Jiao}, H., {Brandenberger}, R., \& {Refregier}, A. 2023, arXiv e-prints,
  arXiv:2304.06429, \dodoi{10.48550/arXiv.2304.06429}

\bibitem[{{Johnson} {et~al.}(2021){Johnson}, {Leja}, {Conroy}, \&
  {Speagle}}]{Johnson2021}
{Johnson}, B.~D., {Leja}, J., {Conroy}, C., \& {Speagle}, J.~S. 2021, \apjs,
  254, 22, \dodoi{10.3847/1538-4365/abef67}

\bibitem[{{Kocevski} {et~al.}(2023){Kocevski}, {Onoue}, {Inayoshi}, {Trump},
  {Arrabal Haro}, {Grazian}, {Dickinson}, {Finkelstein}, {Kartaltepe},
  {Hirschmann}, {Fujimoto}, {Juneau}, {Amorin}, {Bagley}, {Barro}, {Bell},
  {Bisigello}, {Calabro}, {Cleri}, {Cooper}, {Ding}, {Grogin}, {Ho}, {Inoue},
  {Jiang}, {Jones}, {Koekemoer}, {Li}, {Li}, {McGrath}, {Molina}, {Papovich},
  {Perez-Gonzalez}, {Pirzkal}, {Wilkins}, {Yang}, \& {Yung}}]{Kocevski2023}
{Kocevski}, D.~D., {Onoue}, M., {Inayoshi}, K., {et~al.} 2023, arXiv e-prints,
  arXiv:2302.00012, \dodoi{10.48550/arXiv.2302.00012}

\bibitem[{{Kriek} {et~al.}(2006){Kriek}, {van Dokkum}, {Franx}, {Quadri},
  {Gawiser}, {Herrera}, {Illingworth}, {Labb{\'e}}, {Lira}, {Marchesini},
  {Rix}, {Rudnick}, {Taylor}, {Toft}, {Urry}, \& {Wuyts}}]{Kriek2006}
{Kriek}, M., {van Dokkum}, P.~G., {Franx}, M., {et~al.} 2006, \apjl, 649, L71,
  \dodoi{10.1086/508371}

\bibitem[{{Labb{\'e}} {et~al.}(2023{\natexlab{a}}){Labb{\'e}}, {van Dokkum},
  {Nelson}, {Bezanson}, {Suess}, {Leja}, {Brammer}, {Whitaker}, {Mathews},
  {Stefanon}, \& {Wang}}]{Labbe2023}
{Labb{\'e}}, I., {van Dokkum}, P., {Nelson}, E., {et~al.} 2023{\natexlab{a}},
  \nat, 616, 266, \dodoi{10.1038/s41586-023-05786-2}

\bibitem[{{Labb{\'e}} {et~al.}(2023{\natexlab{b}}){Labb{\'e}}, {Greene},
  {Bezanson}, {Fujimoto}, {Furtak}, {Goulding}, {Matthee}, {Naidu}, {Oesch},
  {Atek}, {Brammer}, {Chemerynska}, {Coe}, {Cutler}, {Dayal}, {Feldmann},
  {Franx}, {Glazebrook}, {Leja}, {Marchesini}, {Maseda}, {Nanayakkara},
  {Nelson}, {Pan}, {Papovich}, {Price}, {Suess}, {Wang}, {Whitaker},
  {Williams}, \& {Zitrin}}]{Labbe2023b}
{Labb{\'e}}, I., {Greene}, J.~E., {Bezanson}, R., {et~al.} 2023{\natexlab{b}},
  arXiv e-prints, arXiv:2306.07320, \dodoi{10.48550/arXiv.2306.07320}

\bibitem[{{Leja} {et~al.}(2013){Leja}, {van Dokkum}, \& {Franx}}]{Leja2013}
{Leja}, J., {van Dokkum}, P., \& {Franx}, M. 2013, \apj, 766, 33,
  \dodoi{10.1088/0004-637X/766/1/33}

\bibitem[{{Looser} {et~al.}(2023){Looser}, {D'Eugenio}, {Maiolino}, {Witstok},
  {Sandles}, {Curtis-Lake}, {Chevallard}, {Tacchella}, {Johnson}, {Baker},
  {Suess}, {Carniani}, {Ferruit}, {Arribas}, {Bonaventura}, {Bunker},
  {Cameron}, {Charlot}, {Curti}, {de Graaff}, {Maseda}, {Rawle}, {Rix},
  {Rodriguez Del Pino}, {Smit}, {{\"U}bler}, {Willott}, {Alberts}, {Egami},
  {Eisenstein}, {Endsley}, {Hausen}, {Rieke}, {Robertson}, {Shivaei},
  {Williams}, {Boyett}, {Chen}, {Ji}, {Jones}, {Kumari}, {Nelson}, {Perna},
  {Saxena}, \& {Scholtz}}]{Looser2023}
{Looser}, T.~J., {D'Eugenio}, F., {Maiolino}, R., {et~al.} 2023, arXiv
  e-prints, arXiv:2302.14155, \dodoi{10.48550/arXiv.2302.14155}

\bibitem[{{Marshall} {et~al.}(2022){Marshall}, {Wilkins}, {Di Matteo}, {Roper},
  {Vijayan}, {Ni}, {Feng}, \& {Croft}}]{Marshall2022}
{Marshall}, M.~A., {Wilkins}, S., {Di Matteo}, T., {et~al.} 2022, \mnras, 511,
  5475, \dodoi{10.1093/mnras/stac380}

\bibitem[{{Mason} {et~al.}(2023){Mason}, {Trenti}, \& {Treu}}]{Mason2023}
{Mason}, C.~A., {Trenti}, M., \& {Treu}, T. 2023, \mnras, 521, 497,
  \dodoi{10.1093/mnras/stad035}

\bibitem[{{Matthee} {et~al.}(2023){Matthee}, {Naidu}, {Brammer}, {Chisholm},
  {Eilers}, {Goulding}, {Greene}, {Kashino}, {Labbe}, {Lilly}, {Mackenzie},
  {Oesch}, {Weibel}, {Wuyts}, {Xiao}, {Bordoloi}, {Bouwens}, {van Dokkum},
  {Illingworth}, {Kramarenko}, {Maseda}, {Mason}, {Meyer}, {Nelson}, {Reddy},
  {Shivaei}, {Simcoe}, \& {Yue}}]{Matthee2023}
{Matthee}, J., {Naidu}, R.~P., {Brammer}, G., {et~al.} 2023, arXiv e-prints,
  arXiv:2306.05448, \dodoi{10.48550/arXiv.2306.05448}

\bibitem[{{Menci} {et~al.}(2022){Menci}, {Castellano}, {Santini}, {Merlin},
  {Fontana}, \& {Shankar}}]{Menci2022}
{Menci}, N., {Castellano}, M., {Santini}, P., {et~al.} 2022, \apjl, 938, L5,
  \dodoi{10.3847/2041-8213/ac96e9}

\bibitem[{{Moster} {et~al.}(2020){Moster}, {Naab}, \& {White}}]{Moster2020}
{Moster}, B.~P., {Naab}, T., \& {White}, S. D.~M. 2020, \mnras, 499, 4748,
  \dodoi{10.1093/mnras/staa3019}

\bibitem[{{Mowla} {et~al.}(2019){Mowla}, {van der Wel}, {van Dokkum}, \&
  {Miller}}]{Mowla2019}
{Mowla}, L., {van der Wel}, A., {van Dokkum}, P., \& {Miller}, T.~B. 2019,
  \apjl, 872, L13, \dodoi{10.3847/2041-8213/ab0379}

\bibitem[{{Naab} {et~al.}(2009){Naab}, {Johansson}, \& {Ostriker}}]{Naab2009}
{Naab}, T., {Johansson}, P.~H., \& {Ostriker}, J.~P. 2009, \apjl, 699, L178,
  \dodoi{10.1088/0004-637X/699/2/L178}

\bibitem[{{Naidu} {et~al.}(2022{\natexlab{a}}){Naidu}, {Oesch}, {van Dokkum},
  {Nelson}, {Suess}, {Brammer}, {Whitaker}, {Illingworth}, {Bouwens},
  {Tacchella}, {Matthee}, {Allen}, {Bezanson}, {Conroy}, {Labbe}, {Leja},
  {Leonova}, {Magee}, {Price}, {Setton}, {Strait}, {Stefanon}, {Toft},
  {Weaver}, \& {Weibel}}]{Naidu2022a}
{Naidu}, R.~P., {Oesch}, P.~A., {van Dokkum}, P., {et~al.} 2022{\natexlab{a}},
  \apjl, 940, L14, \dodoi{10.3847/2041-8213/ac9b22}

\bibitem[{{Naidu} {et~al.}(2022{\natexlab{b}}){Naidu}, {Oesch}, {Setton},
  {Matthee}, {Conroy}, {Johnson}, {Weaver}, {Bouwens}, {Brammer}, {Dayal},
  {Illingworth}, {Barrufet}, {Belli}, {Bezanson}, {Bose}, {Heintz}, {Leja},
  {Leonova}, {Marques-Chaves}, {Stefanon}, {Toft}, {van der Wel}, {van Dokkum},
  {Weibel}, \& {Whitaker}}]{Naidu2022b}
{Naidu}, R.~P., {Oesch}, P.~A., {Setton}, D.~J., {et~al.} 2022{\natexlab{b}},
  arXiv e-prints, arXiv:2208.02794, \dodoi{10.48550/arXiv.2208.02794}

\bibitem[{{Oesch} {et~al.}(2016){Oesch}, {Brammer}, {van Dokkum},
  {Illingworth}, {Bouwens}, {Labb{\'e}}, {Franx}, {Momcheva}, {Ashby}, {Fazio},
  {Gonzalez}, {Holden}, {Magee}, {Skelton}, {Smit}, {Spitler}, {Trenti}, \&
  {Willner}}]{Oesch2016}
{Oesch}, P.~A., {Brammer}, G., {van Dokkum}, P.~G., {et~al.} 2016, \apj, 819,
  129, \dodoi{10.3847/0004-637X/819/2/129}

\bibitem[{{Oesch} {et~al.}(2023){Oesch}, {Brammer}, {Naidu}, {Bouwens},
  {Chisholm}, {Illingworth}, {Matthee}, {Nelson}, {Qin}, {Reddy}, {Shapley},
  {Shivaei}, {van Dokkum}, {Weibel}, {Whitaker}, {Wuyts}, {Covelo-Paz},
  {Endsley}, {Fudamoto}, {Giovinazzo}, {Herard-Demanche}, {Kerutt},
  {Kramarenko}, {Labbe}, {Leonova}, {Lin}, {Magee}, {Marchesini}, {Maseda},
  {Mason}, {Matharu}, {Meyer}, {Neufeld}, {Prieto Lyon}, {Schaerer}, {Sharma},
  {Shuntov}, {Smit}, {Stefanon}, {Wyithe}, \& {Xiao}}]{Oesch2023}
{Oesch}, P.~A., {Brammer}, G., {Naidu}, R.~P., {et~al.} 2023, \mnras, 525,
  2864, \dodoi{10.1093/mnras/stad2411}

\bibitem[{{Ono} {et~al.}(2023){Ono}, {Harikane}, {Ouchi}, {Yajima}, {Abe},
  {Isobe}, {Shibuya}, {Wise}, {Zhang}, {Nakajima}, \& {Umeda}}]{Ono2022}
{Ono}, Y., {Harikane}, Y., {Ouchi}, M., {et~al.} 2023, \apj, 951, 72,
  \dodoi{10.3847/1538-4357/acd44a}

\bibitem[{{Onoue} {et~al.}(2023){Onoue}, {Inayoshi}, {Ding}, {Li}, {Li},
  {Molina}, {Inoue}, {Jiang}, \& {Ho}}]{Onoue2023}
{Onoue}, M., {Inayoshi}, K., {Ding}, X., {et~al.} 2023, \apjl, 942, L17,
  \dodoi{10.3847/2041-8213/aca9d3}

\bibitem[{{Papovich} {et~al.}(2011){Papovich}, {Finkelstein}, {Ferguson},
  {Lotz}, \& {Giavalisco}}]{Papovich2011}
{Papovich}, C., {Finkelstein}, S.~L., {Ferguson}, H.~C., {Lotz}, J.~M., \&
  {Giavalisco}, M. 2011, \mnras, 412, 1123,
  \dodoi{10.1111/j.1365-2966.2010.17965.x}

\bibitem[{{Peng} {et~al.}(2002){Peng}, {Ho}, {Impey}, \& {Rix}}]{Peng2002}
{Peng}, C.~Y., {Ho}, L.~C., {Impey}, C.~D., \& {Rix}, H.-W. 2002, \aj, 124,
  266, \dodoi{10.1086/340952}

\bibitem[{{Peng} {et~al.}(2010){Peng}, {Ho}, {Impey}, \& {Rix}}]{Peng2010x}
---. 2010, \aj, 139, 2097, \dodoi{10.1088/0004-6256/139/6/2097}

\bibitem[{{P{\'e}rez-Gonz{\'a}lez} {et~al.}(2023){P{\'e}rez-Gonz{\'a}lez},
  {Barro}, {Annunziatella}, {Costantin}, {Garc{\'\i}a-Argum{\'a}nez},
  {McGrath}, {M{\'e}rida}, {Zavala}, {Arrabal Haro}, {Bagley}, {Backhaus},
  {Behroozi}, {Bell}, {Bisigello}, {Buat}, {Calabr{\`o}}, {Casey}, {Cleri},
  {Coogan}, {Cooper}, {Cooray}, {Dekel}, {Dickinson}, {Elbaz}, {Ferguson},
  {Finkelstein}, {Fontana}, {Franco}, {Gardner}, {Giavalisco},
  {G{\'o}mez-Guijarro}, {Grazian}, {Grogin}, {Guo}, {Huertas-Company}, {Jogee},
  {Kartaltepe}, {Kewley}, {Kirkpatrick}, {Kocevski}, {Koekemoer}, {Long},
  {Lotz}, {Lucas}, {Papovich}, {Pirzkal}, {Ravindranath}, {Somerville},
  {Tacchella}, {Trump}, {Wang}, {Wilkins}, {Wuyts}, {Yang}, \&
  {Yung}}]{PerezGonzalez2023}
{P{\'e}rez-Gonz{\'a}lez}, P.~G., {Barro}, G., {Annunziatella}, M., {et~al.}
  2023, \apjl, 946, L16, \dodoi{10.3847/2041-8213/acb3a5}

\bibitem[{Perrin {et~al.}(2014)Perrin, Sivaramakrishnan, Lajoie, Elliott,
  Pueyo, Ravindranath, \& Albert}]{Perrin2014}
Perrin, M.~D., Sivaramakrishnan, A., Lajoie, C.-P., {et~al.} 2014, in Space
  Telescopes and Instrumentation 2014: Optical, Infrared, and Millimeter Wave,
  Vol. 9143 (SPIE), 91433X, \dodoi{10.1117/12.2056689}

\bibitem[{{Salpeter}(1955)}]{Salpeter1955}
{Salpeter}, E.~E. 1955, \apj, 121, 161, \dodoi{10.1086/145971}

\bibitem[{{Sersic}(1968)}]{Sersic1968}
{Sersic}, J.~L. 1968, {Atlas de Galaxias Australes}

\bibitem[{{Shibuya} {et~al.}(2015){Shibuya}, {Ouchi}, \&
  {Harikane}}]{Shibuya2015}
{Shibuya}, T., {Ouchi}, M., \& {Harikane}, Y. 2015, \apjs, 219, 15,
  \dodoi{10.1088/0067-0049/219/2/15}

\bibitem[{{Stefanon} {et~al.}(2021){Stefanon}, {Bouwens}, {Labb{\'e}},
  {Illingworth}, {Gonzalez}, \& {Oesch}}]{Stefanon2021}
{Stefanon}, M., {Bouwens}, R.~J., {Labb{\'e}}, I., {et~al.} 2021, \apj, 922,
  29, \dodoi{10.3847/1538-4357/ac1bb6}

\bibitem[{{Steinhardt} {et~al.}(2023){Steinhardt}, {Kokorev}, {Rusakov},
  {Garcia}, \& {Sneppen}}]{Steinhardt2022}
{Steinhardt}, C.~L., {Kokorev}, V., {Rusakov}, V., {Garcia}, E., \& {Sneppen},
  A. 2023, \apjl, 951, L40, \dodoi{10.3847/2041-8213/acdef6}

\bibitem[{{Szomoru} {et~al.}(2010){Szomoru}, {Franx}, {van Dokkum}, {Trenti},
  {Illingworth}, {Labb{\'e}}, {Bouwens}, {Oesch}, \& {Carollo}}]{Szomoru2010}
{Szomoru}, D., {Franx}, M., {van Dokkum}, P.~G., {et~al.} 2010, \apjl, 714,
  L244, \dodoi{10.1088/2041-8205/714/2/L244}

\bibitem[{{Tal} {et~al.}(2009){Tal}, {van Dokkum}, {Nelan}, \&
  {Bezanson}}]{Tal2009}
{Tal}, T., {van Dokkum}, P.~G., {Nelan}, J., \& {Bezanson}, R. 2009, \aj, 138,
  1417, \dodoi{10.1088/0004-6256/138/5/1417}

\bibitem[{{Torrey} {et~al.}(2017){Torrey}, {Wellons}, {Ma}, {Hopkins}, \&
  {Vogelsberger}}]{Torrey2017}
{Torrey}, P., {Wellons}, S., {Ma}, C.-P., {Hopkins}, P.~F., \& {Vogelsberger},
  M. 2017, \mnras, 467, 4872, \dodoi{10.1093/mnras/stx370}

\bibitem[{{Trujillo} {et~al.}(2007){Trujillo}, {Conselice}, {Bundy}, {Cooper},
  {Eisenhardt}, \& {Ellis}}]{Trujillo2007}
{Trujillo}, I., {Conselice}, C.~J., {Bundy}, K., {et~al.} 2007, \mnras, 382,
  109, \dodoi{10.1111/j.1365-2966.2007.12388.x}

\bibitem[{{Trujillo} {et~al.}(2011){Trujillo}, {Ferreras}, \& {de La
  Rosa}}]{trujillo2011}
{Trujillo}, I., {Ferreras}, I., \& {de La Rosa}, I.~G. 2011, \mnras, 415, 3903,
  \dodoi{10.1111/j.1365-2966.2011.19017.x}

\bibitem[{{Trujillo} {et~al.}(2006){Trujillo}, {Feulner}, {Goranova}, {Hopp},
  {Longhetti}, {Saracco}, {Bender}, {Braito}, {Della Ceca}, {Drory},
  {Mannucci}, \& {Severgnini}}]{Trujilo2006}
{Trujillo}, I., {Feulner}, G., {Goranova}, Y., {et~al.} 2006, \mnras, 373, L36,
  \dodoi{10.1111/j.1745-3933.2006.00238.x}

\bibitem[{{van der Wel} {et~al.}(2008){van der Wel}, {Holden}, {Zirm}, {Franx},
  {Rettura}, {Illingworth}, \& {Ford}}]{vanderWel2008}
{van der Wel}, A., {Holden}, B.~P., {Zirm}, A.~W., {et~al.} 2008, \apj, 688,
  48, \dodoi{10.1086/592267}

\bibitem[{{van Dokkum} {et~al.}(2008){van Dokkum}, {Franx}, {Kriek}, {Holden},
  {Illingworth}, {Magee}, {Bouwens}, {Marchesini}, {Quadri}, {Rudnick},
  {Taylor}, \& {Toft}}]{Dokkum2008}
{van Dokkum}, P.~G., {Franx}, M., {Kriek}, M., {et~al.} 2008, \apjl, 677, L5,
  \dodoi{10.1086/587874}

\bibitem[{{van Dokkum} {et~al.}(2010){van Dokkum}, {Whitaker}, {Brammer},
  {Franx}, {Kriek}, {Labb{\'e}}, {Marchesini}, {Quadri}, {Bezanson},
  {Illingworth}, {Muzzin}, {Rudnick}, {Tal}, \& {Wake}}]{Dokkum2010}
{van Dokkum}, P.~G., {Whitaker}, K.~E., {Brammer}, G., {et~al.} 2010, \apj,
  709, 1018, \dodoi{10.1088/0004-637X/709/2/1018}

\bibitem[{{van Dokkum} {et~al.}(2015){van Dokkum}, {Nelson}, {Franx}, {Oesch},
  {Momcheva}, {Brammer}, {F{\"o}rster Schreiber}, {Skelton}, {Whitaker}, {van
  der Wel}, {Bezanson}, {Fumagalli}, {Illingworth}, {Kriek}, {Leja}, \&
  {Wuyts}}]{Dokkum2015}
{van Dokkum}, P.~G., {Nelson}, E.~J., {Franx}, M., {et~al.} 2015, \apj, 813,
  23, \dodoi{10.1088/0004-637X/813/1/23}

\bibitem[{{Wang} {et~al.}(2023){Wang}, {Leja}, {Bezanson}, {Johnson},
  {Khullar}, {Labb{\'e}}, {Price}, {Weaver}, \& {Whitaker}}]{Wang2023}
{Wang}, B., {Leja}, J., {Bezanson}, R., {et~al.} 2023, \apjl, 944, L58,
  \dodoi{10.3847/2041-8213/acba99}

\bibitem[{{Weaver} {et~al.}(2023){Weaver}, {Cutler}, {Pan}, {Whitaker},
  {Labbe}, {Price}, {Bezanson}, {Brammer}, {Marchesini}, {Leja}, {Wang},
  {Furtak}, {Zitrin}, {Atek}, {Coe}, {Dayal}, {van Dokkum}, {Feldmann},
  {Forster Schreiber}, {Franx}, {Fujimoto}, {Fudamoto}, {Glazebrook}, {de
  Graaff}, {Greene}, {Juneau}, {Kassin}, {Kriek}, {Khullar}, {Maseda}, {Mowla},
  {Muzzin}, {Nanayakkara}, {Nelson}, {Oesch}, {Pacifici}, {Papovich}, {Setton},
  {Shapley}, {Smit}, {Stefanon}, {Taylor}, {Weibel}, \&
  {Williams}}]{Weaver2023}
{Weaver}, J.~R., {Cutler}, S.~E., {Pan}, R., {et~al.} 2023, arXiv e-prints,
  arXiv:2301.02671, \dodoi{10.48550/arXiv.2301.02671}

\bibitem[{{Williams} {et~al.}(2014){Williams}, {Giavalisco}, {Cassata},
  {Tundo}, {Wiklind}, {Guo}, {Lee}, {Barro}, {Wuyts}, {Bell}, {Conselice},
  {Dekel}, {Faber}, {Ferguson}, {Grogin}, {Hathi}, {Huang}, {Kocevski},
  {Koekemoer}, {Koo}, {Ravindranath}, \& {Salimbeni}}]{Williams2014}
{Williams}, C.~C., {Giavalisco}, M., {Cassata}, P., {et~al.} 2014, \apj, 780,
  1, \dodoi{10.1088/0004-637X/780/1/1}

\bibitem[{{Yang} {et~al.}(2022){Yang}, {Morishita}, {Leethochawalit},
  {Castellano}, {Calabr{\`o}}, {Treu}, {Bonchi}, {Fontana}, {Mason}, {Merlin},
  {Paris}, {Trenti}, {Roberts-Borsani}, {Bradac}, {Vanzella}, {Vulcani},
  {Marchesini}, {Ding}, {Nanayakkara}, {Birrer}, {Glazebrook}, {Jones},
  {Boyett}, {Santini}, {Strait}, \& {Wang}}]{Yang2022}
{Yang}, L., {Morishita}, T., {Leethochawalit}, N., {et~al.} 2022, \apjl, 938,
  L17, \dodoi{10.3847/2041-8213/ac8803}

\end{thebibliography}
\bibliographystyle{aasjournal}
\end{document}